\newif\ifconfver

\newif\ifonecoltab

\newif\ifplainver  
\plainvertrue

\ifplainver
\confverfalse                         
\fi

\ifconfver
\documentclass[10pt,journal]{IEEEtran}
\else
\ifplainver
\documentclass[11pt]{article}
\usepackage{fullpage}
\else
\documentclass[12pt,draftcls,onecolumn]{IEEEtran}
\fi
\fi

\usepackage{calc,amsfonts,amssymb,amsmath,bm,url,color,theorem,graphicx,cite,epstopdf,nicefrac}
\usepackage{psfrag,subfigure,float}
\usepackage[algoruled,linesnumbered]{algorithm2e}
\usepackage{multirow}
\usepackage[normalem]{ulem}
\usepackage{framed,subfigure}
\usepackage{amsfonts, amsmath, color, graphicx,framed,bbm,epstopdf}
\usepackage{amsmath,amssymb,bm,bbold}
\usepackage{stmaryrd}

\newtheorem{Lemma}{Lemma}
\newtheorem{Prop}{Proposition}
\newtheorem{Theorem}{Theorem}
\newtheorem{Def}{Definition}

\theorembodyfont{\rmfamily}

\usepackage{caption}
\usepackage{todonotes}

\newcommand{\Y}{\boldsymbol{Y}}

\newcommand{\X}{\boldsymbol{X}}

\newcommand{\U}{\boldsymbol{U}}

\newcommand{\A}{\boldsymbol{A}}
\newcommand{\B}{\boldsymbol{B}}

\newcommand{\x}{\boldsymbol{x}}

\newcommand{\y}{\boldsymbol{y}}

\newcommand{\tX}{\underline{\bm X}}
\newcommand{\tZ}{\underline{\bm Z}}

\DeclareMathOperator*{\minimize}{\textrm{minimize}}

\usepackage{xcolor}
\usepackage{psfrag,framed}
\usepackage{lipsum}
\usepackage{chapterbib}
\PassOptionsToPackage{normalem}{ulem}
\usepackage{ulem}
\definecolor{shadecolor}{RGB}{220,220,220}

\definecolor{orange}{RGB}{255,107,0}

\begin{document}

%
%
%

\newcommand{\papertitle}{
	\resizebox{\linewidth}{!}{On Recoverability of Randomly Compressed Tensors with Low CP Rank}
}

\newcommand{\paperabstract}{%
Our interest lies in the recoverability properties of compressed tensors under the \textit{canonical polyadic decomposition} (CPD) model. The considered problem is well-motivated in many applications, e.g., hyperspectral image and video compression. Prior work studied this problem under somewhat special assumptions---e.g., the latent factors of the tensor are sparse or drawn from absolutely continuous distributions. We offer an alternative result: We show that if the tensor is compressed by a subgaussian linear mapping, then the tensor is recoverable if the number of measurements is on the same order of magnitude as that of the model parameters---without strong assumptions on the latent factors. Our proof is based on deriving a \textit{restricted isometry property} (R.I.P.) under the CPD model via set covering techniques, and thus exhibits a flavor of classic compressive sensing. The new recoverability result enriches the understanding to the compressed CP tensor recovery problem; it offers theoretical guarantees for recovering tensors whose elements are not necessarily continuous or sparse.
}


\ifplainver

\date{\today}

\title{\papertitle}

\author{
	Shahana Ibrahim$^\dagger$, Xiao Fu$^\dagger$, and Xingguo Li$^\ast$\\
	~\\
	$^\dagger$School of Electrical Engineering and Computer Science\\
	Oregon State Univeristy\\
	Email: (ibrahish,xiao.fu)@oregonstate.edu\\	
    ~\\
	$^\ast$Department of Computer Science\\
	Princeton Univeristy\\
	Email: xingguol@princeton.edu
}

\date{}

\maketitle

\begin{abstract}
	Our interest lies in the recoverability properties of compressed tensors under the \textit{canonical polyadic decomposition} (CPD) model. The considered problem is well-motivated in many applications, e.g., hyperspectral image and video compression. Prior work studied this problem under somewhat special assumptions---e.g., the latent factors of the tensor are sparse or drawn from absolutely continuous distributions. We offer an alternative result: We show that if the tensor is compressed by a subgaussian linear mapping, then the tensor is recoverable if the number of measurements is on the same order of magnitude as that of the model parameters---without strong assumptions on the latent factors. Our proof is based on deriving a \textit{restricted isometry property} (R.I.P.) under the CPD model via set covering techniques, and thus exhibits a flavor of classic compressive sensing. The new recoverability result enriches the understanding to the compressed CP tensor recovery problem; it offers theoretical guarantees for recovering tensors whose elements are not necessarily continuous or sparse.
\end{abstract}

\else
\title{\papertitle}

\ifconfver \else {\linespread{1.1} \rm \fi

	\author{Xiao Fu and Kejun Huang
		
		\thanks{
			%
			
			This work is supported in part by National Science Foundation under Project NSF ECCS-1608961 and ECCS 1808159.
			
			X. Fu is with the School of Electrical Engineering and Computer Science, Oregon State University, Corvallis, OR 97331, United States. email: xiao.fu@oregonstate.edu
			
			K. Huang is the Department of Department of Computer and Information Science and Engineering, University of Florida, Gainesville, FL 32611, United States.  email:kejun.huang@ufl.edu

		}
	}
	
	\maketitle
	
	\ifconfver \else
	\begin{center} \vspace*{-2\baselineskip}
	\end{center}
	\fi
	
	\begin{abstract}
		\paperabstract
	\end{abstract}
	
	\begin{IEEEkeywords}\vspace{-0.0cm}%
		Large-scale tensor decomposition, block-term decomposition, accelerated first-order method, nonconvex optimization
	\end{IEEEkeywords}

	\ifconfver \else \IEEEpeerreviewmaketitle} \fi

\fi

\ifconfver \else
\ifplainver \else
\newpage
\fi \fi

\section{Introduction}
Many signal processing problems boil down to an inverse problem.
Consider a system of linear equations, i.e.,
\begin{equation}\label{eq:le}
  \y =\bm \Phi\bm x
\end{equation}
where $\bm \Phi\in\mathbb{R}^{M\times J}$ denotes a sensing system, 
$\y\in\mathbb{R}^M$ is the observed measurement vector, and $\x\in\mathbb{R}^J$ is the signal of interest.
The task of the inverse probem is to recover $\x$ from $\y$ with the knowledge of the sensing system $\bm \Phi$.
In many cases, the number of measurements is much smaller than that of the signal dimension, i.e., $M\ll J$, which makes the inverse problem highly under-determined. In general, recovering $\x$ is impossible under such cases---an infinite number of solutions exist because $\bm \Phi$ admits a nontrivial null space \cite{GHGolub1996}.

To recover $\x$ when $M\ll J$, one workaround is to exploit some special structure of $\x$. For example, in \textit{compressive sensing} (CS) \cite{baraniuk2008simple,candes2008restricted,foucart2013mathematical}, it is now well-known that if $\x$ is a sparse vector, recovery is possible under some conditions.
This is not entirely surprising, since if the number of nonzero elements in $\x$ is small, the system of linear equations in \eqref{eq:le} is ``essentially over-determined''.
An extension of CS is \textit{low-rank matrix recovery} (LMR) \cite{recht2010guaranteed,candes2011tight}. Similarly, when $\x={\rm vec}(\X)$ and $\X\in\mathbb{R}^{I_1\times I_2}$ is a low-rank matrix, the number of unknowns can be much smaller than $I_1I_2$, which again makes the inverse problem virtually over-determined. Both CS and LMR have received tremendous attention due to their wide spectrum of applications \cite{shen2012unified,zhang2014hyperspectral,zhao2010low,fu2015self}.

As a step further, tensor compression and recovery \cite{wang2017sparse,rauhut2017low,yang2016robust,wang2018hyperspectral,friedland2014compressive} is also quite well-motivated, since many real-world signals are naturally low-rank tensors.
For example, remotely sensed hyperspectral images are third-order tensors (each data entry has two spatial coordinates and one spectral coordinate) \cite{bioucas2012hyperspectral,Ma2013}.
For sensing devices deployed on satellites or aircrafts,  compression is needed for transmitting the acquired data back to earth stations \cite{wang2017sparse,wang2018hyperspectral}. This way, the communication overhead can be substantially reduced.
A lot of data arising in machine learning are also tensors, e.g., social network data \cite{papalexakis2017tensors} and traffic flow data \cite{yang2016robust}. Compressing such data helps save space for storage and overhead for transmission.

A number of works have considered recoverability properties in tensor compression. The recent work \cite{rauhut2017low} considers recovering tensors from random measurements under the Tucker model, hierarchical Tucker (HT) model, and the tensor train (TT) model, respectively.
The works in \cite{sidiropoulos2012multi,sidiropoulos2014parallel,bousse2018linear} consider recovering compressed tensors under the {\it canonical polyadic decomposition} (CPD) model.
Notably, \cite{sidiropoulos2012multi} shows that tensors with low CP rank and sparse latent factors can be recovered from compressed measurements, via solving a series of CS problems in the latent domain.
The work \cite{bousse2018linear} shows that if the latent factors are drawn from a certain joint continuous distribution, then the compressed tensor can be recovered almost surely if the number of measurements is larger than or equal to that of the parameters in the CPD model.
These are all plausible results---showing that recovering compressed tensors is viable under some conditions.

In this work, we offer a new result regarding recoverability of compressed tensors that follow the CPD model (or, CP tensors for short). 
Our result is different from the existing recoverability arguments in \cite{sidiropoulos2012multi,bousse2018linear} in the sense that no sparsity or distributional assumption is imposed on the latent factors in our case.
Our technical approach is based on set covering and deriving a new \textit{restricted isometry property} (R.I.P.) for CP tensors, which is similar to the route of proof in \cite{rauhut2017low} that considers the Tucker, HT and TT models. Showing that a compression system satisfies R.I.P. for CP tensors is challenging since the latent factors of the CPD model cannot be orthogonalized in most cases---as CPD is essentially unique under mild conditions. However, orthogonality of the latent factors are hinged on to show R.I.P. for Tucker/HT/TT tensors.
Nevertheless, we show that recovering a tensor with low CP rank from limited measurements is possible---if the latent factors are reasonably well-conditioned.
Unlike existing results, our recovery proof does not impose sparsity or continuity constraints on the latent factors of the CP tensor, and thus covers cases whose recoverability properties were unknown before.

\section{Problem Statement and Background}
\subsection{Tensor Preliminaries}
An $N$th order tensor $\tX\in\mathbb{R}^{I_1\times \ldots \times I_N}$ is an array whose elements are indexed by $N$ indices, namely, $i_1,\ldots,i_N$, which can be considered as a high-dimensional extension of a matrix. 
Unlike matrices whose definition for rank is singular, there are many different definitions of tensor rank \cite{kolda2009tensor,sidiropoulos2017tensor}. Among them, a popular and useful one is CP rank.
 Specifically, the CP rank of a tensor $\tX$, ${\sf rank}_{\rm C}(\tX)=F$ means that $F$ is the smallest integer such that $\tX$ is expressed as follows:
\begin{align} 
	\tX = \sum_{f=1}^F \A_{(1)}(:,f)\circ \ldots \circ \A_{(N)}(:,f)\quad {\in \mathbb{R}^{I_1\times \ldots \times I_N}}, \label{eq:tensorX}
\end{align}
where $\A_{(n)}\in\mathbb{R}^{I_n\times F}$ denotes the mode-$n$ latent factor under CPD and ``$\circ$'' is the outer product; see details in \cite{sidiropoulos2017tensor}. The term $\A_{(1)}(:,f)\circ \ldots \circ \A_{(N)}(:,f)$ is called a \textit{rank-one tensor}.
CPD is seemingly similar to the matrix SVD, since SVD can also be understood as a summation of rank-one matrices.
However, the $\A_{(n)}$'s in \eqref{eq:tensorX} cannot always be orthogonalized as in the SVD case, because the CPD is essentially unique under mild conditions; see details in the tutorial on CPD uniqueness \cite{sidiropoulos2017tensor}.


Besides CPD, many other tensor decomposition models exist in the literature. For example, Tucker decomposition \cite{tucker1966some}, hierarchical Tucker (HT) decomposition \cite{rauhut2017low} and tensor train (TT) decomposition\cite{ose2011tensortrain} are also considered useful in representing tensor data in parsimonious ways. 

\subsection{The Compressed Tensor Recovery Problem}
Our interest lies in the following linear system:
\begin{align}     \y ={\cal A}\left(\tX_\natural\right),           \label{eq:y}    
\end{align}
where $\tX_\natural$ is the ``ground-truth signal'' of interest, ${\cal A}(\cdot):\mathbb{R}^{I_1\times \ldots \times I_N}\rightarrow \mathbb{R}^M$ is a linear mapping, i.e., ${\cal A}(\tX_\natural)= \bm \Phi{\rm vec}(\tX_\natural)$ where $\bm \Phi\in\mathbb{R}^{M\times (\prod_{n=1}^NI_n)}$.
When $M\ll \prod_{n=1}^N I_n$, the inverse problem of recovering $\tX_\natural$ from $\y$ may have an infinite number of solutions.
However, if $\tX_\natural$ is a low CP rank tensor with
${\sf rank}_{\rm C}(\tX_\natural)= F$
and the number of linear measurements $M$ is larger than the number of unknown parameters (i.e., $(I_1+\ldots + I_N-1)F$), then the inverse problem is ``essentially over-determined'', and recovering $\tX_\natural$ is possible---which is the starting point of our work.

Consider a recovery criterion as follows:

\vspace{.15cm}
\noindent
\boxed{
\begin{minipage}{.95\linewidth}
{\bf Recovery Criterion:}
	\begin{subequations}\label{eq:criterion}
	\begin{align}
	\minimize_{\tX}&~\left\|\y -{\cal A}(\tX) \right\|_{\rm F}^2\\
	{\rm subject~to}&~{\sf rank}_{\rm C}(\tX)\leq F	
	\end{align}
	\end{subequations}
\end{minipage}
}
\vspace{.15cm} 

\noindent
We are concerned with the \textit{recoverability} properties of Criterion~\eqref{eq:criterion}. Specifically, assume that one can solve Problem~\eqref{eq:criterion} to optimality using a certain algorithm, does the optimal solution(s) (denoted by $\tX_{\rm opt}$) recover the uncompressed signal $\tX_\natural$ under some conditions on ${\cal A}(\cdot)$ and $\tX_\natural$? In addition, how many measurements are needed to recover $\tX_\natural$?
\vspace{-0.5em}
\subsection{Related Work}
\subsubsection{Tucker, HT, and TT Tensors} The recent work in \cite{rauhut2017low} considered a similar problem but the tensors admit low-rank Tucker, HT, or TT representation.
Assuming that a \textit{subgaussian} mapping ${\cal A}(\cdot)$ is used, then when the number of measurements is on the same order of magnitude as that of the tensor parameters, then recovery is possible under the Tucker, HT, and TT models.


\subsubsection{CP Tensors} It is also of great interest to study the recoverability properties of CP tensors, since exact CPD exists for every tensor without modeling error \cite{sidiropoulos2017tensor}. In addition, the CP representation is very economical in terms of the number of unknowns (i.e. $\sum_{n=1}^N I_NF-F$), which only increases linearly with the tensor order $N$ (while Tucker's number of parameters increases exponentially with $N$).

Several notable works on CP tensor recovery appeared in recent years. Specifically, the work in \cite{sidiropoulos2012multi} considers a case where $\tX_{\natural}$'s latent factors are all sparse. 
Using a special sensing system $\bm \Phi=\bm \Phi_1\otimes \ldots \otimes \bm \Phi_N$ where ``$\otimes$'' denotes the Kronecker product,
the tensor recovery problem can be recast as a series classical CS problems in the latent factor domain---which helps establish the identifiability of $\A_{(n)}$'s, thereby that of $\tX_{\natural}$.
The work in \cite{sidiropoulos2014parallel} extends this latent factor recovery-based approach to dense $\A_{(n)}$'s, with the price of using many more different compressed measurements in parallel. The works in \cite{sidiropoulos2012multi,sidiropoulos2014parallel} are both based on the assumption that the compressed measurements are small tensors that admit unique CPD.
In \cite{bousse2018linear}, this assumption is relaxed. There, almost sure recoverability of $\tX_{\natural}$ is shown under the assumption that $\A_{(n)}$'s and $\bm \Phi$ are drawn from certain joint continuous distributions. The sample complexity proved in \cite{bousse2018linear} is appealing, which is exactly the number of unknowns.
The caveat is that the $\A_{(n)}$'s have to follow a certain continuous distribution---which means that some important types of tensors (e.g., tensors with discrete latent factors that have applications in machine learning \cite{papalexakis2017tensors,yang2017learning,fu2015joint}) may not be covered by the recoverability theorem in \cite{bousse2018linear}.
\vspace{-0.2em}
\section{Main Result}
In this work, we consider the recoverability problem for CP tensors as in \cite{sidiropoulos2012multi,sidiropoulos2014parallel,bousse2018linear}.
Unlike these prior works, we do not restrict $\tX_\natural$ and its compressed versions to admit unique CPD or assume that $\tX_\natural$'s latent factors are drawn from joint continuous distributions.
As a trade-off, we restrict the entries of the sensing matrix $\bm \Phi$ to be zero-mean i.i.d. \textit{subgaussian} (see \cite{dirksen2016dimensionality} for more details about subgaussian matrices).
Subgaussian sensing matrices are widely used in compressive sensing and dimensionality reduction, since they have a lot of appealing features \cite{dirksen2016dimensionality,baraniuk2008simple,recht2010guaranteed,rauhut2017low}. Fortunately, in many scenarios, the sensing/compressing matrix is under control of the system designers (e.g., in communications)---and thus assuming subguassianity of $\bm \Phi$ is considered reasonable in such cases.
\subsection{Recoverability  under CP Tensor R.I.P.}
Let us consider the following definition:
\vspace{-1.2em}
\begin{Def}(CP tensor R.I.P.)
	Assume that for all $\tX\in \{\tX\in\mathbb{R}^{I_1\times \ldots \times I_N}~|~{\sf rank}_{\rm C}(\tX)\leq F$ and $0\leq\delta_F<1\}$, the following holds:
	\begin{equation}\label{eq:tensorRIP}
	(1-\delta_F)\|\tX\|_{\rm F}^2 \leq \|{\cal A}(\tX)\|_{\rm F}^2 \leq (1+\delta_F)\|\tX\|_{\rm F}^2.
	\end{equation}
	Then, it is said that the mapping ${\cal A}(\cdot)$ satisfies the restricted isometry property (R.I.P.) with parameter $\delta_{F}$ for tensors with CP rank being $F$.
\end{Def}
\vspace{-1.2em}
If a mapping on a set of tensors satisfies R.I.P., then recoverability of this set of tensors can be readily established: 
\vspace{-1em}
\begin{Lemma}\label{lem:rip} (Recoverability under R.I.P.)
	If ${\cal A}(\cdot)$ satisfies R.I.P. for tensors whose CP rank is smaller than or equal to $2F$ with parameter $0\leq \delta_{2F}<1$. Assume that ${\sf rank}_{\rm C}(\tX_\natural)\leq F$. Then, the optimal solution to Problem~\eqref{eq:criterion} is $\tX_{\rm opt} = \tX_{\natural}$.
\end{Lemma}

{\it Proof}:
The proof is the same as that in matrix recovery \cite{recht2010guaranteed}. Assume that there is a rank-$F$ tensor $\tZ$ and $\tZ\neq \tX_\natural$ which satisfies
${\cal A}(\tZ)=\y.$
Then, we have
\[ \bm  0=\|{\cal A}(\tX_\natural-\tZ)\|_{\rm F}^2 \geq (1-\delta_{2F})\|\tX_\natural-\tZ\|_{\rm F}^2>0, \]
which is a contradiction. 
In the above, we have used the facts that ${\sf rank}_{\rm C}(\tX_\natural-\tZ)\leq 2F$ and that ${\cal A}(\cdot)$ satisfies R.I.P. for all rank-$2F$ CP tensors. \hfill $\square$

From Lemma~\ref{lem:rip}, one can see that, if we could prove that for all the tensors in ${\cal S}=\{\tX~ |~ {\sf rank}_{\rm C}(\tX)\leq 2F,~\tX\in\mathbb{R}^{I_1\times\ldots\times I_N} \}$, some ${\cal A}(\cdot)$ satisfies R.I.P. with $0\leq\delta_{2F}<1$, then the recoverability can be established.
Showing this for all rank-$2F$ tensors is, unfortunately, challenging. As we will see, it turns out that the conditioning of $\A_{(n)}$'s plays an important role of establishing R.I.P. for low-rank CP tensors. This is quite different from the low-rank matrix (or the Tucker/HT/TT tensor) case, where only the matrix size and rank matter.
This contrast makes sense, since the matrix latent factors under SVD are always orthonormal, and thus the condition numbers of the latent factors are constants. But for CP tensors, since $\A_{(n)}$'s are essentially unique and not orthogonalizable in many cases, the impact of their conditioning naturally shows up.
To proceed, we define the following parameter:
\vspace{-1em}
\begin{Def}
The condition number $\kappa(\tX)$ of the CP tensor $\tX\in\mathbb{R}^{I_1\times \ldots \times I_N}$ is defined as follows:
\[\kappa(\tX) =\frac{\prod_{n=1}^N \sigma_{\max}(\A_{(n)}) }{\sigma_{\min}\left(  \odot_{n=1}^N \A_{(n)} \right)}.\]
\end{Def}

One can see that $\kappa(\tX)<\infty$ implies that ${\rm rank}(\A_{(1)}\odot \ldots \odot \A_{(N)})=F$, which is a \textit{necessary condition} for the CPD of $\tX$ being essentially unique \cite{sidiropoulos2017tensor}.
The parameter $\kappa(\tX) $ is clearly related to the condition numbers of $\A_{(n)}$'s. 
This may be clearer when $I_n\geq F$ and ${\rm rank}(\A_{(n)})=F$ for all $n$. Under such cases, we have
\begin{subequations}
	\begin{align}
	\sigma_{\min}\left(  \odot_{n=1}^N \A_{(n)} \right)  & = \min_{\|\x\|_2=1}~\left\|\left(  \odot_{n=1}^N \A_{(n)} \right)\x\right\|_2\\
	&=  \min_{\|\x\|_2=1}~\left\|\left(  \otimes_{n=1}^N \A_{(n)} \right){\bm P}\x\right\|_2\\
	& \geq   \prod_{n=1}^N\sigma_{\min}(\A_{(n)})\|\bm P\|_2\|\x\|_2\\
	& = \prod_{n=1}^N\sigma_{\min}(\A_{(n)})
	\end{align}
\end{subequations}
where $\odot$ denotes the Khatri-Rao product and $\bm P$ is a column selection matrix (and thus $\|\bm P\|_2=\sigma_{\rm max}(\bm P)=\sigma_{\min}(\bm P)=1$) and we have used the fact that the columns of $\A\odot\B$ is a subset of the columns of $\A\otimes\B$.
The above leads to
$ \kappa(\tX) \leq  \frac{\prod_{n=1}^N \sigma_{\max}(\A_{(n)}) }{\prod_{n=1}^N\sigma_{\min}(\A_{(n)})}=\prod_{n=1}^N{\rm cond}(\A_{(n)}), $
where ${\rm cond}(\Y)$ denotes the matrix condition number of $\Y$.
Hence, $\kappa(\tX)$ can be understood as a parameter that reflects the conditioning of the latent factors. From the above, another note is that $\prod_{n=1}^N \sigma_{\max}(\A_{(n)})  \geq \sigma_{\max}(\odot_{n=1}^N \A_{(n)})$, resulting in 
$ \kappa(\tX) \geq 1$, which resembles the property of the matrix condition number, i.e., ${\rm cond}(\bm Y)\geq 1$ for any $\bm Y$.


With this parameter defined, our main result is stated in the following theorem:
\begin{Theorem}\label{thm:main}
	Assume that ${\cal A}(\cdot):\mathbb{R}^{I_1\times\ldots\times I_N}\rightarrow \mathbb{R}^M$ is a mapping such that ${\cal A}(\tX_\natural)= \bm \Phi{\rm vec}(\tX_\natural)$, where $\bm \Phi\in\mathbb{R}^{M\times (\prod_{n=1}^NI_n)}$ has i.i.d. zero-mean $\sqrt{\alpha/M}$-subgaussian entries. In addition, assume that $  \kappa(\tX_\natural)\leq \tau $ and ${\sf rank}_{\rm C}(\tX_\natural)\leq F$.
	Then, for a certain constant $C>0$, the criterion in \eqref{eq:criterion} recovers $\tX_\natural$ at its optimal solution with a probability larger than or equal to $1-\eta$ if
	\begin{equation*}
	M> C\alpha^2\max \left \{ \left(1+2\sum_{n=1}^N I_nF\right)\log(3(N+1)\tau),\log\left(\eta^{-1}\right)  \right\}.
	\end{equation*}	
\end{Theorem}
Note that $\sqrt{\alpha/M}$ is known as the subgaussian parameter which is related to the subgaussian distribution that generates the entries of $\bm \Phi$.
For example, ${\cal N}(0,\frac{\alpha}{M})$ is $\sqrt{\alpha/M}$-subgaussian; see more details in \cite{dirksen2016dimensionality,wainwright2019highdimensional}. 
Also note that since we only need $\delta_{2F}<1$ to establish recoverability, the lower bound of $M$ does not contain $\delta_{2F}$ explicitly.

From Theorem~\ref{thm:main}, one can see that with $M=\Omega(\sum_{n=1}^N I_nF)$, recovering $\tX_\natural$ from compressed measurements is possible.
This $M$ and the number of unknowns $(I_1+\ldots + I_N-1)F$ have the same order of magnitude---which is quite plausible.
In addition, there is no sparsity or continuous distribution assumptions on $\A_{(n)}$, which means that Theorem~\ref{thm:main} may be able to cover cases where previous recoverability results in \cite{sidiropoulos2012multi,bousse2018linear} do not support.

\subsection{Proof of Theorem 1}
In this section, we outline the proof of Theorem~\ref{thm:main} in a concise way. Some of the details can be found in the supplementary materials.
Consider the following set of low-rank tensors:
\begin{equation*}
\begin{aligned}
&{\cal S}_{F,\tau} = \left\{  \widetilde{\tX}~\left|~\widetilde{\tX}=\frac{\tX}{\|\tX\|_{\rm F}},{\sf rank}_{\rm C}(\tX)\leq F,~\kappa(\tX)\leq \tau \right.\right\}.
\end{aligned}
\end{equation*}
We will show that \eqref{eq:tensorRIP} holds for $\widetilde{\tX}\in{\cal S}_{F,\tau}$ with high probability if $\bm \Phi_{ij}$ is drawn from a subgaussian distribution. This will imply that \eqref{eq:tensorRIP} holds for all the $\tX$'s associated with $\widetilde{\tX}\in{\cal S}_{F,\tau}$ since the mapping in \eqref{eq:tensorRIP} is linear.
Note that 
\begin{align*}
\frac{\tX(i_1,\ldots,i_n)}{\|\tX\|_{\rm F}} &= \sum_{f=1}^F\underbrace{ \frac{\prod_{n=1}^N \sigma_{\max}(\A_{(n)}) }{\|\tX\|_{\rm F}}}_{\tilde{\lambda}}\prod_{n=1}^N \underbrace{\frac{ \A_{(n)}(i_n,f)}{  \|\A_{(n)}\|_2} }_{\tilde{\A}_{(n)}(i_n,f)}\\
& = \sum_{f=1}^F \tilde{\lambda} \prod_{n=1}^N \tilde{\A}_{(n)}(i_n,f).
\end{align*}
Since
$\|\tX\|_{\rm F} =\|(\odot_{n=1}^N \A_{(n)}){\bm 1}\|_2 \geq \sigma_{\min}\left(\odot_{n=1}^N \A_{(n)} \right)\sqrt{F}$, 
\begin{align*}
\tilde{\lambda} &= \frac{\prod_{n=1}^N \sigma_{\max}(\A_{(n)}) }{\|\tX\|_{\rm F}} \le \frac{\prod_{n=1}^N \sigma_{\max}(\A_{(n)}) }{\sigma_{\min}\left(  \odot_{n=1}^N \A_{(n)} \right)\sqrt{F}}  = \frac{\kappa(\tX)}{\sqrt{F}} .
\end{align*}
Consequently, we have
\begin{align}\label{eq:prop}
\sigma_{\rm max}(\tilde{\A}_{(n)})=\|\tilde{\A}_{(n)}\|_2 = 1,\quad{\lambda} \leq \nicefrac{ \kappa(\tX)}{\sqrt{F}} \leq \nicefrac{\tau}{\sqrt{F}}.
\end{align}
For notational simplicity, we now represent all the tensors in ${\cal S}_{F,\tau}$ as
${\tX}=  \left\llbracket \lambda, {\A}_{(1)},\ldots,{\A}_{(N)} \right\rrbracket \in {\cal S}_{F,\tau}$, where $\A_{(n)}$ and $\lambda$ satisfy \eqref{eq:prop} and $\|\tX\|_{\rm F}=1$.
The set ${\cal S}_{F,\tau} $ has an infinite number of elements. To establish R.I.P. we construct an $\varepsilon$-net (w.r.t. Euclidean norm) that covers ${\cal S}_{F,\tau}$.
An $\varepsilon$-net of ${\cal S}_{F,\tau}$, denoted as $\bar{\cal S}_{F,\tau}$, is a finite set such that for any $\tX\in{\cal S}_{F,\tau}$, one can find a $\bar{\tX}\in\bar{\cal S}_{F,\tau}$
such that
$           \|\bar{\tX}-\tX\|_{\rm F}\leq \varepsilon      $ \cite{vershynin2010introduction}.
We have the following proposition:

\begin{Prop}\label{prop:cover}
	There exists an $\varepsilon$-net $\bar{\cal S}_{F,\tau}\subseteq {\cal S}_{F,\tau}$ with respect to the Fronenius norm such that the cardinality of $\bar{\cal S}_{F,\tau}$ is upper bounded by the following inequality:
	\[  |\bar{\cal S}_{F,\tau}|\leq(3(N+1)\tau/\varepsilon)^{1+\sum_{n=1}^N I_nF} .\]
\end{Prop}
The proof of Proposition~\ref{prop:cover} is given in the supplementary materials. With Proposition~\ref{prop:cover} at hands, we show that
\begin{Prop}\label{prop:RIP}
	For $\tX\in {\cal S}_{F,\tau}$ and zero-mean $\sqrt{\alpha/M}$-subgaussian sensing matrix $\bm \Phi\in\mathbb{R}^{M\times \prod_{n=1}^N}I_n$, $\delta_F$-R.I.P holds for $0\leq \delta_F<1$ and for a certain constant $C>0$ with probability larger than $1-\eta$ provided that
	\begin{equation*}
M\geq C\alpha^2\delta_F^{-2}\max \left \{ \left(1+\sum_{n=1}^N I_nF\right)\log(3(N+1)\tau),\log\left(\eta^{-1}\right)  \right\}.
	\end{equation*}
\end{Prop}
Proposition \ref{prop:RIP} invokes Corollary 5.4 in \cite{dirksen2016dimensionality} (see more details of the proof in the supplementary materials).

Combining Propositions~\ref{prop:cover}-\ref{prop:RIP} and Lemma~\ref{lem:rip}, and the fact that recoverability holds with $\delta_{2F}<1$,
one can easily show Theorem~\ref{thm:main}.

\section{Numerical Validation}
In this section, we present numerical results to validate Theorem \ref{thm:main}. We randomly generate the latent factors of third-order tensors with CP rank $F$ such that the condition number of the latent factors satisfy ${\rm cond}(\bm A_{(1)})={\rm cond}(\bm A_{(2)})={\rm cond}(\bm A_{(3)})=\widetilde{\kappa}$. In order to generate the latent factor with desired condition number, we first generate the entries of the latent factors uniformly at random. Then we change the singular values of the latent factor, while keeping the singular vectors unchanged. 

Using the latent factors, we generate the tensor $\tX$ using Eq.~\eqref{eq:tensorX}. This way, we have generated $\widetilde{\tX} =\tX/\|\tX\|_{\rm F}  \in \mathcal{S}_{F, \tau}$ where $\tau = \widetilde{\kappa}^3$. We employ $\bm \Phi \in \mathbb{R}^{M \times (I_1+I_2+I_3)F}$ such that the entries are randomly chosen from the normal distribution with zero mean and variance $\frac{1}{M}$. This makes the entries of $\bm \Phi$  to be i.i.d zero mean $\sqrt{1/M}$-subgaussian. The observations $\bm y \in \mathbb{R}^M$ are then generated using Eq.~\eqref{eq:y}. In order to solve the tensor recovery problem in \eqref{eq:criterion}, we employ the Gauss-Newton based algorithm proposed in \cite{bousse2018linear}. We stop the algorithm when the relative change in the objective function is less than the machine accuracy. For each $\widetilde{\kappa}$, we run 100 random trials and each trial is counted towards a successful tensor recovery if the \textit{mean squared error} (MSE) is lower than $10^{-10}$, where the MSE is defined as
$
\text{MSE} = \|{\tX}-{\tX}_{\text{rec}}\|^2_{\rm F}/(I_1I_2I_3)
$
where ${\tX}_{\text{rec}}$ is the recovered tensor.

\begin{figure}[h]
	\centering
		\captionsetup{justification=centering}
	\includegraphics[scale=0.45]{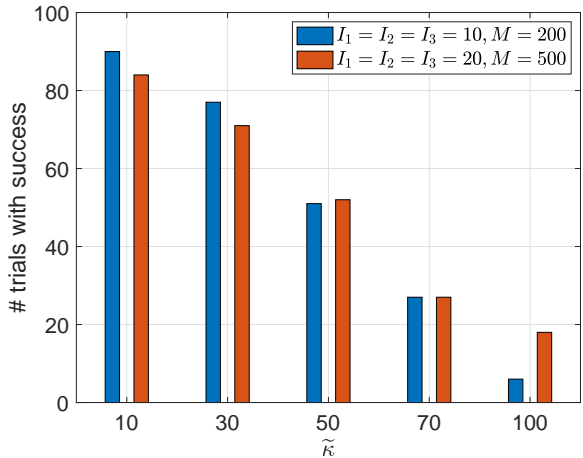}
	\caption{Effect of $\widetilde{\kappa}$ on successful tensor recovery.} 
	\label{fig:rec}
	\vspace{-.5cm}
\end{figure}

Fig.~\ref{fig:rec} shows the number of successful recovery cases against the condition number $\widetilde{\kappa}$ for different values of $I_n$ and $F=3$. 
Note that the tensor recovery problem in \eqref{eq:criterion} is NP-hard \cite{hillar2013most} and thus numerical optimizers may not necessarily output optimal solutions. However, when $F$ is small, the problem can be solved very well by the algorithm in \cite{bousse2018linear} according to our extensive simulations. The numerical results here can therefore serve as reasonably reliable references. 
It can be observed that as the condition number $\tilde{\kappa}$ increases, successful tensor recovery becomes harder to attain under both settings. This is consistent with the result in Theorem~\ref{thm:main} which indicates that a larger $\tau$ needs larger $M$ for successful tensor recovery. 
\vspace{-1em}

\section{Conclusion}
In this work, we considered the recoverability problem for compressed CP tensors.
Unlike previous works which tackled this problem leveraging CPD uniqueness of the compressed tensors or assumptions on the latent factors' distribution, we offered a recoverability theory without making such assumptions. The result derived in this work can potentially cover more cases in practice. The proof also offers insights on how the conditioning of the latent factors of a CPD model can affect recoverability of compressed tensors. We also presented experimental results supporting our theoretical claims.

\bibliographystyle{IEEEtran}

\begin{thebibliography}{10}
\providecommand{\url}[1]{#1}
\csname url@samestyle\endcsname
\providecommand{\newblock}{\relax}
\providecommand{\bibinfo}[2]{#2}
\providecommand{\BIBentrySTDinterwordspacing}{\spaceskip=0pt\relax}
\providecommand{\BIBentryALTinterwordstretchfactor}{4}
\providecommand{\BIBentryALTinterwordspacing}{\spaceskip=\fontdimen2\font plus
\BIBentryALTinterwordstretchfactor\fontdimen3\font minus
  \fontdimen4\font\relax}
\providecommand{\BIBforeignlanguage}[2]{{%
\expandafter\ifx\csname l@#1\endcsname\relax
\typeout{** WARNING: IEEEtran.bst: No hyphenation pattern has been}%
\typeout{** loaded for the language `#1'. Using the pattern for}%
\typeout{** the default language instead.}%
\else
\language=\csname l@#1\endcsname
\fi
#2}}
\providecommand{\BIBdecl}{\relax}
\BIBdecl

\bibitem{GHGolub1996}
G.~H. Golub and C.~F.~V. Loan, \emph{Matrix Computations}.\hskip 1em plus 0.5em
  minus 0.4em\relax The Johns Hopkins University Press, 1996.

\bibitem{baraniuk2008simple}
R.~Baraniuk, M.~Davenport, R.~DeVore, and M.~Wakin, ``A simple proof of the
  restricted isometry property for random matrices,'' \emph{Constructive
  Approximation}, vol.~28, no.~3, pp. 253--263, 2008.

\bibitem{candes2008restricted}
E.~J. Candes, ``The restricted isometry property and its implications for
  compressed sensing,'' \emph{Comptes rendus mathematique}, vol. 346, no. 9-10,
  pp. 589--592, 2008.

\bibitem{foucart2013mathematical}
S.~Foucart and H.~Rauhut, \emph{A mathematical introduction to compressive
  sensing}.\hskip 1em plus 0.5em minus 0.4em\relax Birkh{\"a}user Basel, 2013,
  vol.~1, no.~3.

\bibitem{recht2010guaranteed}
B.~Recht, M.~Fazel, and P.~A. Parrilo, ``Guaranteed minimum-rank solutions of
  linear matrix equations via nuclear norm minimization,'' \emph{SIAM review},
  vol.~52, no.~3, pp. 471--501, 2010.

\bibitem{candes2011tight}
E.~J. Candes and Y.~Plan, ``Tight oracle inequalities for low-rank matrix
  recovery from a minimal number of noisy random measurements,'' \emph{IEEE
  Trans. Inf. Theory}, vol.~57, no.~4, pp. 2342--2359, 2011.

\bibitem{shen2012unified}
X.~Shen and Y.~Wu, ``A unified approach to salient object detection via low
  rank matrix recovery,'' in \emph{Proc. CVPR 2012}.\hskip 1em plus 0.5em minus
  0.4em\relax IEEE, 2012, pp. 853--860.

\bibitem{zhang2014hyperspectral}
H.~Zhang, W.~He, L.~Zhang, H.~Shen, and Q.~Yuan, ``Hyperspectral image
  restoration using low-rank matrix recovery,'' \emph{IEEE Trans. Geosci.
  Remote Sens.}, vol.~52, no.~8, pp. 4729--4743, 2014.

\bibitem{zhao2010low}
B.~Zhao, J.~P. Haldar, C.~Brinegar, and Z.-P. Liang, ``Low rank matrix recovery
  for real-time cardiac {MRI},'' in \emph{Proc. 2010 IEEE International
  Symposium on Biomedical Imaging: From Nano to Macro}, 2010, pp. 996--999.

\bibitem{fu2015self}
X.~Fu, W.-K. Ma, T.-H. Chan, and J.~M. Bioucas-Dias, ``Self-dictionary sparse
  regression for hyperspectral unmixing: Greedy pursuit and pure pixel search
  are related,'' \emph{IEEE J. Sel. Topics Signal Process.}, vol.~9, no.~6, pp.
  1128--1141, 2015.

\bibitem{wang2017sparse}
Y.~Wang, D.~Meng, and M.~Yuan, ``Sparse recovery: from vectors to tensors,''
  \emph{National Science Review}, 2017.

\bibitem{rauhut2017low}
H.~Rauhut, R.~Schneider, and {\v{Z}}.~Stojanac, ``Low rank tensor recovery via
  iterative hard thresholding,'' \emph{Linear Algebra and its Applications},
  vol. 523, pp. 220--262, 2017.

\bibitem{yang2016robust}
Y.~Yang, Y.~Feng, and J.~A.~K. Suykens, ``Robust low-rank tensor recovery with
  regularized redescending {M}-estimator,'' \emph{IEEE Trans. Neural Net.
  Learning Sys.}, vol.~27, no.~9, pp. 1933--1946, Sept 2016.

\bibitem{wang2018hyperspectral}
Y.~Wang, J.~Peng, Q.~Zhao, Y.~Leung, X.-L. Zhao, and D.~Meng, ``Hyperspectral
  image restoration via total variation regularized low-rank tensor
  decomposition,'' \emph{IEEE J. Sel. Topics Appl. Earth Observ.}, vol.~11,
  no.~4, pp. 1227--1243, 2018.

\bibitem{friedland2014compressive}
S.~Friedland, Q.~Li, and D.~Schonfeld, ``Compressive sensing of sparse
  tensors.'' \emph{IEEE Trans. Image Process.}, vol.~23, no.~10, pp.
  4438--4447, 2014.

\bibitem{bioucas2012hyperspectral}
J.~M. Bioucas-Dias, A.~Plaza, N.~Dobigeon, M.~Parente, Q.~Du, P.~Gader, and
  J.~Chanussot, ``Hyperspectral unmixing overview: Geometrical, statistical,
  and sparse regression-based approaches,'' \emph{IEEE J. Sel. Topics Appl.
  Earth Observ.}

\bibitem{Ma2013}
W.-K. Ma, J.~Bioucas-Dias, T.-H. Chan, N.~Gillis, P.~Gader, A.~Plaza,
  A.~Ambikapathi, and C.-Y. Chi, ``A signal processing perspective on
  hyperspectral unmixing,'' \emph{IEEE Signal Process. Mag.}, vol.~31, no.~1,
  pp. 67--81, Jan 2014.

\bibitem{papalexakis2017tensors}
E.~E. Papalexakis, C.~Faloutsos, and N.~D. Sidiropoulos, ``Tensors for data
  mining and data fusion: Models, applications, and scalable algorithms,''
  \emph{ACM Transactions on Intelligent Systems and Technology (TIST)}, vol.~8,
  no.~2, p.~16, 2017.

\bibitem{sidiropoulos2012multi}
N.~D. Sidiropoulos and A.~Kyrillidis, ``Multi-way compressed sensing for sparse
  low-rank tensors,'' \emph{IEEE Signal Process. Lett.}, vol.~19, no.~11, pp.
  757--760, 2012.

\bibitem{sidiropoulos2014parallel}
N.~D. Sidiropoulos, E.~E. Papalexakis, and C.~Faloutsos, ``Parallel randomly
  compressed cubes: A scalable distributed architecture for big tensor
  decomposition,'' \emph{IEEE Signal Process. Mag.}, vol.~31, no.~5, pp.
  57--70, 2014.

\bibitem{bousse2018linear}
M.~Bouss\'e, N.~Vervliet, I.~Domanov, O.~Debals, and L.~De~Lathauwer, ``Linear
  systems with a canonical polyadic decomposition constrained solution:
  Algorithms and applications,'' \emph{Numerical Linear Algebra with
  Applications}, vol.~25, 2018.

\bibitem{kolda2009tensor}
T.~G. Kolda and B.~W. Bader, ``Tensor decompositions and applications,''
  \emph{SIAM review}, vol.~51, no.~3, pp. 455--500, 2009.

\bibitem{sidiropoulos2017tensor}
N.~D. Sidiropoulos, L.~De~Lathauwer, X.~Fu, K.~Huang, E.~E. Papalexakis, and
  C.~Faloutsos, ``Tensor decomposition for signal processing and machine
  learning,'' \emph{IEEE Trans. Signal Process.}, vol.~65, no.~13, pp.
  3551--3582.

\bibitem{tucker1966some}
L.~R. Tucker, ``Some mathematical notes on three-mode factor analysis,''
  \emph{Psychometrika}, vol.~31, no.~3, pp. 279--311, 1966.

\bibitem{ose2011tensortrain}
I.~V. Oseledets, ``Tensor-train decomposition,'' \emph{SIAM Journal on
  Scientific Computing}, vol.~33, no.~5, pp. 2295--2317, 2011.

\bibitem{yang2017learning}
B.~Yang, X.~Fu, and N.~D. Sidiropoulos, ``Learning from hidden traits: Joint
  factor analysis and latent clustering,'' \emph{IEEE Transactions on Signal
  Processing}, vol.~65, no.~1, pp. 256--269, 2017.

\bibitem{fu2015joint}
X.~Fu, K.~Huang, W.-K. Ma, N.~Sidiropoulos, and R.~Bro, ``Joint tensor
  factorization and outlying slab suppression with applications,'' \emph{IEEE
  Trans. Signal Process.}, vol.~63, no.~23, pp. 6315--6328, 2015.

\bibitem{dirksen2016dimensionality}
S.~Dirksen, ``Dimensionality reduction with subgaussian matrices: a unified
  theory,'' \emph{Foundations of Computational Mathematics}, vol.~16, no.~5,
  pp. 1367--1396, 2016.

\bibitem{wainwright2019highdimensional}
M.~J. Wainwright, \emph{High-Dimensional Statistics: A Non-Asymptotic
  Viewpoint}, ser. Cambridge Series in Statistical and Probabilistic
  Mathematics.\hskip 1em plus 0.5em minus 0.4em\relax Cambridge University
  Press, 2019.

\bibitem{vershynin2010introduction}
R.~Vershynin, \emph{Introduction to the non-asymptotic analysis of random
  matrices}.\hskip 1em plus 0.5em minus 0.4em\relax Cambridge University Press,
  2012, p. 210–268.

\bibitem{hillar2013most}
C.~J. Hillar and L.-H. Lim, ``Most tensor problems are np-hard,'' \emph{Journal
  of the ACM (JACM)}, vol.~60, no.~6, p.~45, 2013.

\end{thebibliography}


\newpage
\vspace{-1em}
\appendix
\section{Proof of Proposition 1} \label{app:prop1}
To proceed, we first show the following lemma:
\vspace{-1em}
\begin{Lemma}\label{eq:mat_2_norm}
	Suppose that $\|{\A}_{(n)}\|_2=\sigma_{\max}(\A_{(n)}) =1,~\forall n \in \{1,\dots,N\}$. For any integer $L\in\mathbb{Z}_+$ such that $L \le N$ and any subset of $\{1,\ldots,N\}$ with $L$ elements, i.e., $\{k_1,\dots,k_L\} \subseteq \{1,\dots,N\}$, we  consider a term $\bm W =\A_{(k_1)}\odot \ldots \odot \A_{(k_\ell)}\odot \bm U \odot  \A_{(k_{\ell+1})}\odot \ldots \odot \A_{(k_L)}$, where $\ell\in\{0,\ldots,L\}$. 
	By definition, when $\ell=0$, $\bm U$ appears in the leftmost position of the term; when $\ell=L$, $\bm U$ is in the rightmost of $\bm W$.
	Then, we have the following:
\[   \|\bm W\|_2 \leq \|\bm U\|_2.  \]
\end{Lemma}
{\it Proof}:
We prove the lemma with $\ell=0$. For $\ell=1,\ldots,L$, the proof is almost identical.

Note that $ \|\U\odot{\A}_{(k_1)}\odot \ldots \odot{\A}_{(k_{L})}\|_2 = \|\U\otimes ({\A}_{(k_1)}\odot \ldots \odot{\A}_{(k_{L})}))\bm P \|_2  $, where $\bm P$ is a submatrix of the identity matrix which does column selection. We have the following chain of inequalities:
	\begin{align*}
&\|\U\odot{\A}_{(k_1)}\odot \ldots \odot{\A}_{(k_{L})}\|_2 \\
&= \|\U\otimes ({\A}_{(k_1)}\odot \ldots \odot{\A}_{(k_{L})}))\bm P \|_2  \\
	&\leq \|\U\|_2 \|({\A}_{(k_1)}\odot \ldots \odot{\A}_{(k_{L})})\|_2 \|\bm P\|_2\\
	&= \|\U\|_2 \|({\A}_{(k_1)}\otimes\ldots \otimes{\A}_{(k_L)})\bm P^{'}\|_2 \|\bm P\|_2\\
	&\leq \|\U\|_2 \|({\A}_{(k_1)}\otimes\ldots \otimes{\A}_{(k_L)})\|_2\|\bm P^{'}\|_2 \|\bm P\|_2\\
	&=  \|\U\|_2 \|{\A}_{(k_1)}\|_2\ldots\|{\A}_{(k_L)}\|_2\|\bm P'\|_2\|\bm P\|_2\leq \|\U\|_2.
	\end{align*}
	where $\bm P'$ is also a proper column selection matrix, and we have used $\|\bm P\|_2=\|\bm P'\|_2=1$.	Note that the last equality holds due to the fact that $\|{\A}_{(k_1)}\otimes\ldots \otimes{\A}_{(k_L)}\|_2 =\|{\A}_{(k_1)}\|_2\ldots\|{\A}_{(k_L)}\|_2.$ \hfill $\square$

Consider a tensor ${\underline{\X}}= \left\llbracket {\lambda}, {\A}_{(1)},\ldots,{\A}_{(N)} \right\rrbracket \in{\cal S}_{F,\tau}$. This tensor can be represented as ${\underline{\X}}  = \sum_{f=1}^F \lambda \left(\circ_{n=1}^N\A_{(n)}(:,f)\right)$, which is a short-hand notation for the expression in \eqref{eq:tensorX}.
Now consider another tensor $\bar{\tX}=  \left\llbracket \bar{\lambda}, \bar{\A}_{(1)},\ldots,\bar{\A}_{(N)} \right\rrbracket \in{\cal S}_{F,\tau}$. 
The Euclidean distance between the two tensors are bounded because of the following inequalities:
\begin{align}
&\|\bar{\underline{\X}}-\underline{\X}\|_{\rm F}\nonumber \\
& =\left\lVert \sum_{f=1}^F\bar{\lambda}\left(\circ_{n=1}^N\bar{\A}_{(n)}(:,f)\right) - \sum_{f=1}^F \lambda\left(\circ_{n=1}^N\A_{(n)}(:,f)\right)\right\rVert_{\rm F} \nonumber \\
&\leq \left\| \sum_{f=1}^F(\bar{\lambda}-\lambda) \left(\circ_{n=1}^N\bar{\A}_{(n)}(:,f)\right)\right\|_{\rm F} \nonumber\\
&\quad+\underbrace{\left\lVert \sum_{f=1}^F{\lambda}\left(\circ_{n=1}^N\bar{\A}_{(n)}(:,f)\right) - \sum_{f=1}^F \lambda\left(\circ_{n=1}^N\A_{(n)}(:,f)\right)\right\rVert_{\rm F}}_{Q_{\lambda}} \nonumber \\
&\leq  \left\|\bar{\bm A}_{(N)}\odot\ldots\odot\bar{\bm A}_{(1)} \right\|_{2} |\bar{\lambda}-\lambda|\sqrt{F}+ Q_{\lambda} \nonumber\\
& \leq |\bar{\lambda}-\lambda|\sqrt{F} + Q_{\lambda}. \label{eq:ABC}
\end{align}
where Eq.~\eqref{eq:ABC} is obtained by Lemma~\ref{eq:mat_2_norm}. Now consider,
\begin{align}
Q_{\lambda} &= \left\lVert \sum_{f=1}^F{\lambda}\left(\circ_{n=1}^N\bar{\A}_{(n)}(:,f)\right) - \sum_{f=1}^F \lambda\left(\circ_{n=1}^N\A_{(n)}(:,f)\right)\right\rVert_{\rm F}\nonumber\\
&\leq \left\|   \sum_{f=1}^F\lambda\left(\bar{\bm A}_{(1)}(:,f) - \A_{(1)}(:,f)\right)\circ\left(\circ_{n=2}^N\bar{\A}_{(n)}(:,f) \right) \right\|_{\rm F} \nonumber \\
&\quad +\underbrace{\left\|  \sum_{f=1}^F\lambda \A_{(1)}(:,f)\circ\left(\circ_{n=2}^N\bar{\A}_{(n)}(:,f)  -  \circ_{n=2}^N{\A}_{(n)}(:,f) \right)\right\|_{\rm F}}_{Q_{\bm A_{(1)}}} \nonumber \\
& \leq  \left\|\bar{\bm A}_{(N)}\odot\ldots\odot\left(\bar{\bm A}_{(1)}-{\bm A}_{(1)}\right) \right\|_{2}| \lambda|\sqrt{F} + Q_{\bm A_{(1)}}\nonumber \\
&  \leq  \|\bar{\A}_{(1)}-\A_{(1)}\|_2 \tau+Q_{\bm A_{(1)}}. \label{eq:ABC_Ftau}
\end{align}
where Eq.~\eqref{eq:ABC_Ftau} is obtained by invoking Lemma~\ref{eq:mat_2_norm} and using the fact $\lambda\sqrt{F} \le \tau$ as given by Eq. \eqref{eq:prop}.
In this way, we can obtain similar inequalities for all $Q_{\bm A_{(n)}},~n=1,\dots,N$ and we can finally establish the below relationship: \\
\vspace{-1em}
\begin{align*}
&\|\tX -\bar{\tX}\|_{\rm F}\leq \sum_{n=1}^N\|\bar{\A}_{(n)}-\A_{(n)}\|_2\tau+ |\bar{\lambda}-\lambda|\sqrt{F}.
\end{align*}

Hence, to show that there exists a $\tau\epsilon$-net covering ${\cal S}_{F,\tau}$
we only need to show that there exists a set covering $\A_{(n)}$ with width $\epsilon/(N+1)$ and the same applies to $ \lambda\sqrt{F}/\tau$.
Since both $\A_{(n)}$ and $ \lambda\sqrt{F}/\tau$ live in respective unit norm balls (unit matrix $2$-norm ball for $\A_{(n)}$ in particular), it is well-known that there exist $\epsilon/(N+1)$-nets that cover them, which have the cardinalities bounded by
$  (3(N+1)/\epsilon)^{I_nF}$ and $ 3(N+1)/\epsilon$
respectively\cite{dirksen2016dimensionality,baraniuk2008simple,recht2010guaranteed}.
Overall, the $\tau\epsilon$-net of ${\cal S}_{F,\tau}$ has $(3(N+1)/\epsilon)^{1+\sum_{n=1}^N I_nF}$ points inside. 
Or, if we let $\varepsilon = \tau\epsilon$, we have $\varepsilon$-net of ${\cal S}_{F,\tau}$ with $(3(N+1)\tau/\varepsilon)^{1+\sum_{n=1}^N I_nF}$ elements.
\section{Proof of Proposition \ref{prop:RIP}} \label{app:prop2}
Consider the following lemma:
\begin{Lemma}[Corollary 5.4 \cite{dirksen2016dimensionality}] \label{lem:covering}
	Let $S_1,\ldots,S_k$ be subsets of a Hilbert space $\mathcal{H}$ and let $S=\cup_{i=1}^k S_i$. Set 
	$ S_{i,nv}=\{ \x/\|\x\|_{\mathcal{H}}:~\x\in S_i  \}$ where $\|.\|_{\mathcal{H}}$ is the induced norm on $\mathcal{H}$ and $S_{nv} =\cup_{i=1}^k S_{i,nv}$.
	Suppose that $S_{i,nv}$ has covering dimension $K_i$ with parameter $c_i$ and base covering $N_{0,i}$ with respect to the induced metric on $\mathcal{H}$. Set $K=\max_i K_i$,  $ c= \max_i c_i$ and $N_0=\max_i~N_{0,i}$. Let $\bm \Phi $ be a subgaussian map which maps $S_{nv}$ to $\mathbb{R}^{M}$. Then, for some constant $C > 0$, for any $0<\delta,\eta<1$, restricted isometry
	constant $\delta_{S,\bm \Phi}$ of $\bm \Phi$ on $S$ satisfies ${\sf Pr}(\delta_{S,\bm \Phi}\geq \delta)\leq \eta$ provided that
	\[  M\geq C\alpha^2\delta^{-2}\max\{ \log k+\log N_0+K\log(c),\log(\eta^{-1})  \}. \]
\end{Lemma}
In our case, $k=1$. The set $\mathcal{S}_{F,\tau}$ belongs to the Hilbert space which is the Euclidean space with Euclidean distance as the induced metric.
If the covering number of a Hilbert space with respect to unit norm ball  is bounded by an expression of the form $N_0 (\frac{c}{\varepsilon})^{K}$ for any $0 < \varepsilon \le 1$,  then $N_0$ is the base covering and $K$ is the covering dimension with parameter $c$ (Def. 5.1, \cite{dirksen2016dimensionality}). Therefore, according to Proposition \ref{prop:cover}, we have $N_0=1$, $ K =  1+\sum_{n=1}^N I_nF$ and $c=3(N+1)\tau.$ By applying these parameters in Lemma~\ref{lem:covering}, we get the result in Proposition \ref{prop:RIP}.

%
%
%

\end{document}